\documentclass[preprint,12pt]{elsarticle}

\usepackage{lineno}

\usepackage{fullpage}

\usepackage[parfill]{parskip}
\usepackage{graphicx}

\usepackage{color}

\journal{Journal of Water Process Engineering}

\newcommand{\inch}{$^{\prime\prime}$}

\begin{document}

\begin{frontmatter}

\title{The Water Purification System for the\\Daya Bay Reactor Neutrino Experiment}
\author[Temple]{J. Wilhelmi}
\author[RPI]{R. Bopp}
\author[BNL]{R. Brown}
\author[Wisc]{J. Cherwinka}
\author[Siena]{J. Cummings}
\author[BNL]{E. Dale}
\author[BNL]{M. Diwan}
\author[RPI]{J. Goett}
\author[BNL]{R.W. Hackenburg}
\author[RPI]{J. Kilduff}
\author[BNL]{L. Littenberg}
\author[SJTU]{G.S. Li}
\author[IHEP]{X.N. Li}
\author[IHEP]{J.C. Liu}
\author[IHEP]{H.Q. Lu}
\author[Temple]{J. Napolitano}
\author[BNL]{C. Pearson}
\author[RPI]{N. Raper}
\author[BNL]{R. Rosero}
\author[RPI]{P. Stoler}
\author[Wisc]{Q. Xiao}
\author[IHEP]{C.G. Yang}
\author[IHEP]{Y. Yang}
\author[BNL]{M. Yeh}
\address[Temple]{Temple University, Philadelphia, PA, USA}
\address[Siena]{Siena College, Loudonville, NY, USA}
\address[RPI]{Rensselaer Polytechnic Institute, Troy, NY, USA}
\address[BNL]{Brookhaven National Laboratory, Upton, NY, USA}
\address[IHEP]{Institute of High Energy Physics, Beijing, China}
\address[SJTU]{Shanghai Jiao Tong University, Shanghai, China}
\address[Wisc]{University of Wisconsin, Madison, WI, USA}
\date{\today}

\begin{abstract}
We describe the design, installation, and operation of a purification system that is able to provide large volumes of high purity ASTM (D1193-91) Type-I water to a high energy physics experiment. The water environment is underground in a lightly sealed system, and this provides significant challenges to maintaining high purity in the storage pools, each of which contains several thousand cubic meters. High purity is dictated by the need for large optical absorption length, which is critical for the operation of the experiment. The system is largely successful, and the water clarity criteria are met. We also include a discussion of lessons learned.
\end{abstract}


\end{frontmatter}

\section{Introduction}
The Daya Bay Reactor Neutrino Experiment~\cite{DayaBay:2012aa,An:2012eh,An:2012bu,An:2013zwz} is a high energy particle physics experiment which aims to measure properties of electron anti-neutrinos (or, for the rest of this paper, simply ``neutrinos'') produced by the Daya Bay Nuclear Power Plant complex located in Southeastern China. One major challenge to neutrino detection is their extremely low probability of interaction requiring large detectors, in this case, gadolinium doped liquid scintillator (Gd-LS) Antineutrino Detectors (ADs).  Despite the high neutrino flux in this experiment, there is major potential interference from a number of sources capable of producing ``non-neutrino'' signals in the Gd-LS. 

For example, the neutrino signal rate is $\sim0.01$~Hz in the detectors located nearest the reactors, and $\sim0.001$~Hz in the detectors farthest away. On the other hand, signals due to radioactivity are $\sim200$~Hz (reduced from $\sim100$~kHz by the presence of the water pools described in this paper), and the cosmic ray flux is $\sim1$~Hz$/$m$^2$ in the near detector halls, and  $\sim0.04$~Hz$/$m$^2$ in the far halls. Clearly these ``background'' sources need to be highly suppressed.

To help solve these problems, we immerse the ADs in large pools of water, as part of a comprehensive muon detection system~\cite{An:2014xx}. As described below, this approach tackles backgrounds both from natural radioactivity and penetrating cosmic rays. This paper focuses on the relationship between water purity and the background suppression performance of the pools.

Natural radioactivity sources include uranium and thorium series radioisotopes in the surrounding rocks, as well as $^{40}$K.  Each AD has at least 2.5~m of water between it and the walls providing sufficient passive shielding against any rock-associated radioactivity.  Radon, a radioactive inert gas arising from U and Th series decay, can find its way into the water volume, and this leads to additional constraints on the design of the pools and the water purification system.

The other important background, directly related to the clarity of water in the pools around the ADs, is cosmic radiation. Energetic cosmic ray collisions in the upper atmosphere lead to ``showers'' of muons and neutrons at the Earth's surface. The ADs are located under hundreds of meters of rock to greatly reduce the flux of cosmogenic neutrons. However, muons are charged particles and produce Cherenkov light while passing through the water. This light is monitored by arrays of 300 to 400 photomultiplier tubes (PMTs) providing a veto signal to the ADs so that cosmic rays are not confused with neutrino events.

It is critical that the water be clear enough so that enough Cherenkov light reaches the PMTs. The water pool PMTs are sensitive to light with wavelengths $\sim300$~nm to $\sim650$~nm, with a peak efficiency at $\sim400$~nm, and Cherenkov light has a $1/($wavelength)$^2$ dependence. The DYB water pools are large, each ten meters deep, two of which measure $\sim10\times16$~m$^2$ and the third measuring $\sim16\times16$~m$^2$. Therefore, the experiment specified that the attenuation length be more than 30~m for these wavelengths. This is well within the absorption length of ultra pure water~\cite{ISI:A1997YF88600025,ISI:A1997YF88600024}, so our goal was to produce the highest purity water possible and maintain it at as high a level as possible, within the constraints of this environment.

\begin{figure}[t]
\centering
\includegraphics[width=\textwidth]{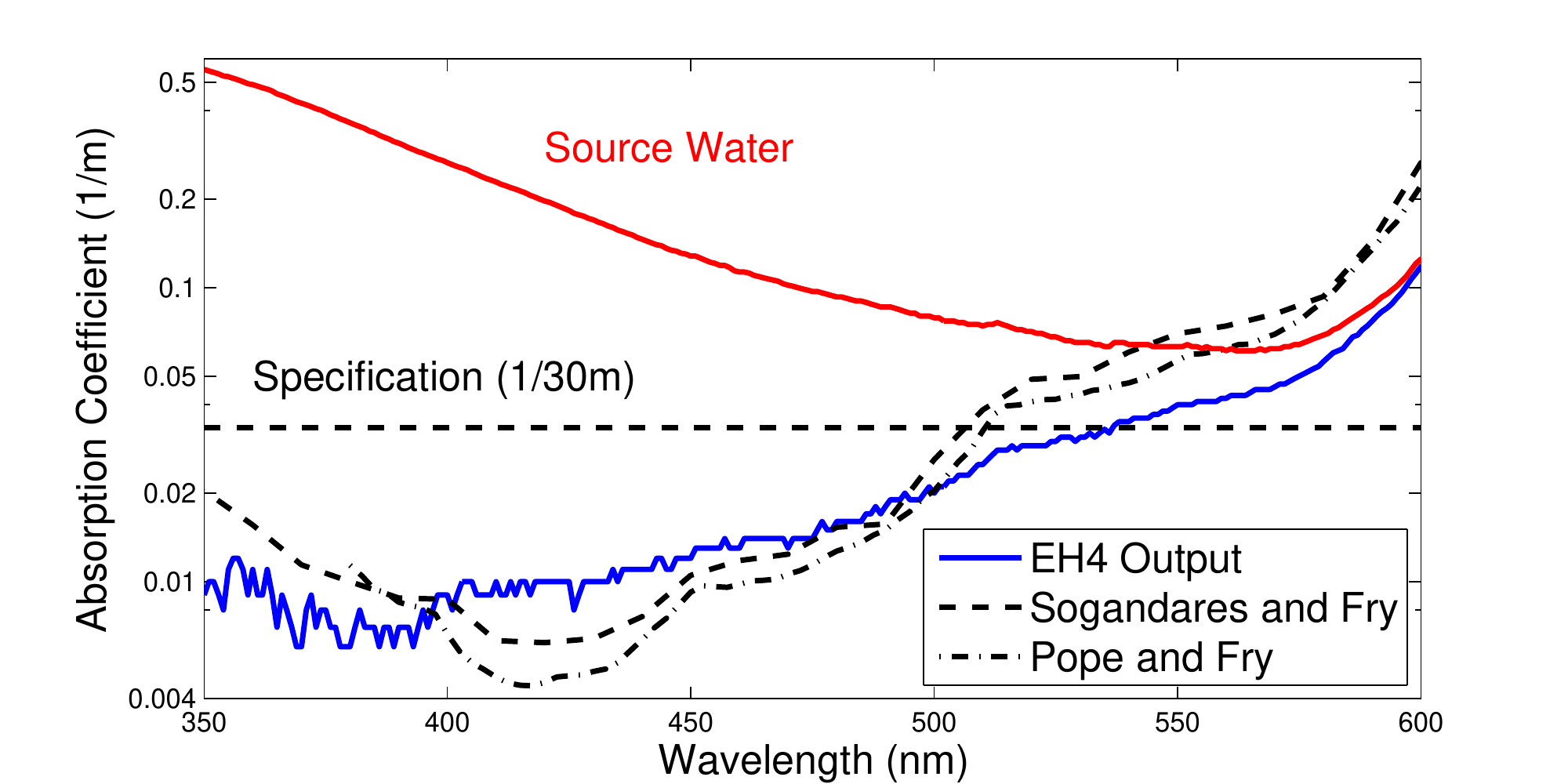}
\caption{Absorption coefficient (reciprocal of attenuation length) of water as a function of wavelength, for DYB source water and for water treated by the Fill/MakeUp stage (EH4). We also show published measurements~\cite{ISI:A1997YF88600025,ISI:A1997YF88600024} of ultra high purity water, and indicate the 30~m attenuation length specified by the neutrino physics experiment, in the wavelength range 400-500~nm.
\label{fig:Absorption}}
\end{figure}

Figure~\ref{fig:Absorption}
illustrates the challenge we faced, and the extent to which we met that challenge. It plots (on a logarithmic vertical scale) the absorption coefficient (i.e. inverse attenuation length) as a function of wavelength for water under different circumstances. Transparency measurements for ultra pure water~\cite{ISI:A1997YF88600025,ISI:A1997YF88600024} are compared to our own analysis of samples of source water and output from our primary purification stage. Our measurements were made with a Shimadzu spectrophotometer, model UV-1800, with a 10~cm long optical cell. An analysis of the municipal source water indicates a turbidity of 0.31~NTU, a pH of 7.06, and a conductivity (resitivity) of 88.4~$\mu$S/cm ($1.13\times10^{-2}$~M$\Omega$-cm). Although the source water at Daya Bay, which comes from an open man-made reservoir  $\sim$2~km away, is well out of specification, our purification system produces water that exceeds the requirements of the Daya Bay experiment.
%Our analysis of the source water contaminants in listed in Table~\ref{tab:SourceWater}.
%\begin{table}
%\begin{center}
%\caption{Source water at Daya Bay August 27, 2007}
%\label{tab:SourceWater}
%\begin{tabular}{|lrl||lrl|}
%\hline
%Contaminant &  Quantity & Unit  & Contaminant &  Quantity & Unit   \\      
%\hline
%Tubidity &0.31 &NTU  & Chromium& $<$0.004 & mg/L \\
%pH &7.06& & Magnesium & 0.46 & mg/L \\
%Conductivity& 88.4 & $\mu$S/cm  & Calcium& 1.25 & mg/L \\
%Hardness &                           7.82 & mg/L  & Potassium & 0.98 & mg/L \\
%CO$_3$$^{-2}$      &                               27.9& mg/L  & Sodium & 15.85 & mg/L \\
%HCO$_3$$^-$        &                            28.3 & mg/L  & Cadmium & $<$0.001 & mg/L \\
%Chlorine & 0.00 & mg/L   & Lead & $<$0.001 & mg/L \\
%Chlorine compounds            &13.6 &mg/L  & &&\\
%Total dissolved solids            & 44 &mg/L     & &&\\
%Residue after evaporation &    36& mg/L  & &&\\
%\hline
%\end{tabular}
%\end{center}
%\end{table}

\begin{figure}
\begin{center}
%\includegraphics[width=0.49\textwidth]{EH3Cutaway.pdf}
%\hfill
%\includegraphics[width=0.49\textwidth]{EH3Pool.pdf}
\includegraphics[width=\textwidth]{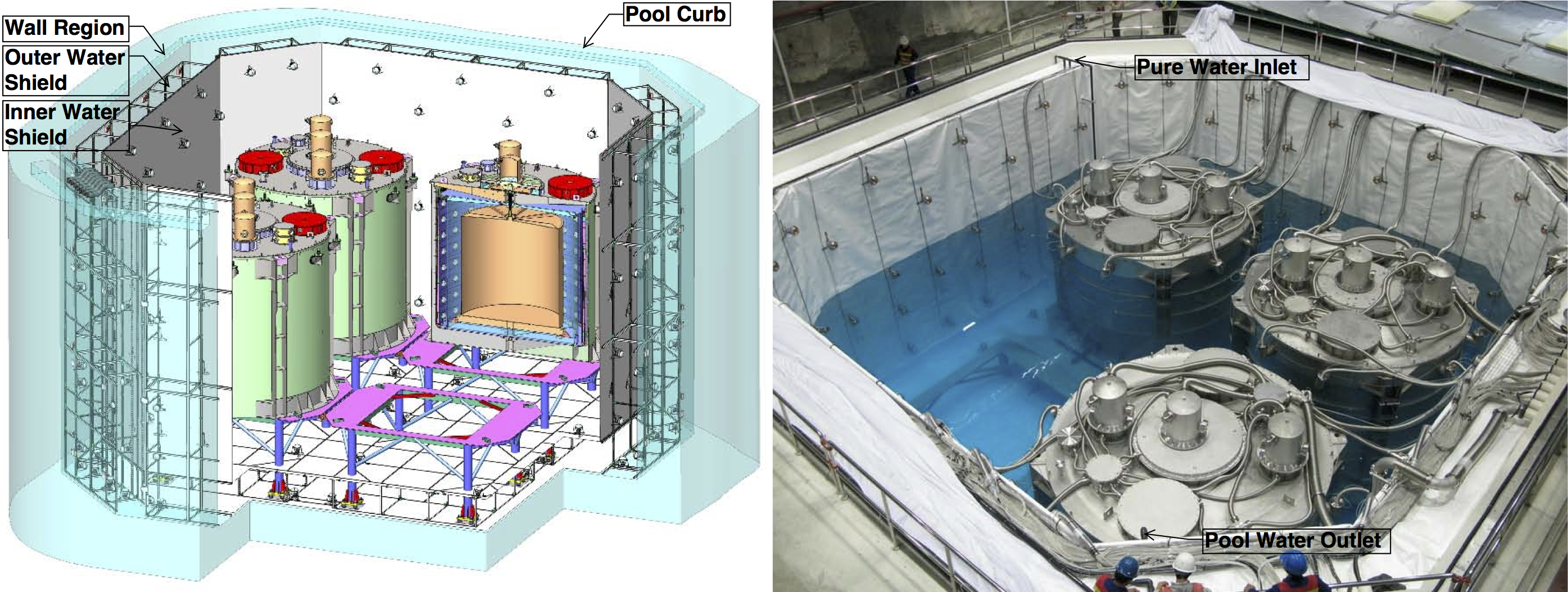}
\caption{Cutaway schematic and photograph of the large ($16\times16\times10$~m$^3$) experimental hall. One of the four antineutrino detectors is left out in each, and the photograph shows the pool partially filled with water. Cherenkov light from cosmic ray muons is detected by the PMTs seen along the sides of the pool. There is in fact a one-meter thick ``outer pool'', within the same excavated cavity, but optically separated from the ``inner pool'' with vertical sheets of Tyvek\textsuperscript{\textregistered}.
\label{fig:Pools}}
\end{center}
\end{figure}

Another challenge arises from the experiment environment, an open pool formed by blasting granite in an underground cavern with personnel and equipment routinely present.
Figure~\ref{fig:Pools}
shows the experimental configuration in the large hall.
When filled with water, the pools are covered with an opaque rubberized sheet, supported by a positive pressure layer of dry N$_2$ gas, to keep out light and atmosphere. Given this environment, it is difficult to limit contaminants that affect water purity. For example, water resistivity at 25$^\circ$C can drop from its theoretical  maximum of 18.18~M$\Omega$-cm to 17.5~M$\Omega$-cm by dissolving 1.0 $\mu$g/L of NaCl~\cite{Morash1994}, so cleanliness of preparation and constant filtration are necessities.

The remainder of this paper describes the design and performance of the water purification system. We begin with our design goals, including a comparison to similar systems used in other high energy physics experiments. We then outline our specific system design for meeting these goals, including a discussion of the monitoring hardware and software, and system maintenance preparations. Next we show performance results, from both startup through steady operation. Finally, we discuss the overall performance and suggest modifications we would pursue for a next generation system.

\section{Water quality requirements and design goals}

Ideally, one would design a water purification system based on allowed concentrations of ion species which meet the optical absorption specifications. Data is in fact
available~\cite{ISI:000200888800002,ISI:A1954UK62900051,ISI:A1957WB82000001,
Ravisankar1988,ISI:A1997XR27100038,Sullivan2006}
on absorption in water with different dissolved salts, and it indicates~\cite{Ravisankar1988} that the optical degradation is tolerable with significant amounts of what might be common ions. However, we concluded that it was too difficult to anticipate what impurities might be present, and to what concentrations, so our decision was to build a system which was prepared to remove {\em all} ions, leaving us with ultra pure water.

Consequently, existing systems from other high energy physics experiments~\cite{Adam:2004fq,Becker-Szendy:1992hr,Boger:1999bb,Fukuda:2002uc,Suekane:2006qi} provided criteria to guide the design of our system. These are summarized in Table~\ref{tab:PreviousExpts}, along with a comparison to parameters for DYB.
\begin{table}
\begin{center}
\caption{Parameters of water purification systems used in previous high energy physics experiments, compared to the DYB near and far pools.
\label{tab:PreviousExpts}}
\begin{tabular}{|lrrrr|}
\hline
&                  &          &         & Hydraulic\\
&  Volume    & \multicolumn{2}{c}{Flow Rate}  & Residence \\
Experiment  &  (m$^3$) & (m$^3$/hr) &(gpm) & (days)\\
\hline
BaBar/DIRC~\cite{Adam:2004fq} & 6 & 0.9 & 4 & 0.28\\
IMB~\cite{Becker-Szendy:1992hr,FosterIMB} & 8000 & 18 & 80 & 18\\
SNO~\cite{Boger:1999bb, Dekok} & 1700 & 9 & 40 & 7.8\\
Super-K~\cite{Fukuda:2002uc} & 50,000 & 30 & 130 & 71\\
KamLAND~\cite{Suekane:2006qi} & 3200 & 8 & 35 & 16.8\\
Milagro~\cite{Atkins:1999gb} & $\approx$4000 & 43 & 190 & 3.9\\
\hline
DYB Far Pool     & 1996 & 9 & 40 & 8.3\\
DYB Near Pools & 1232 & 5 & 22 & 6.7\\
\hline
\end{tabular}
\end{center}
\end{table}

Based on our site specific constraints, and a comparison to the experiments listed in Table~\ref{tab:PreviousExpts}, we settled on the following design criteria for water resident in the DYB pools:
Particulate Size $\leq1.0~\mu$m;
Total Dissolved Solids (TDS)  $\leq4$~ppm;
Resistivity  $\geq15$~M$\Omega$-cm;
Dissolved Oxygen  $\leq0.1$~mg/l; and
Hydraulic Residence Time  $\leq7$~days and 11~days, respectively, for the small and large pools.

Naturally occurring $^{40}$K and isotopes of uranium and thorium make the surrounding granite radioactive. (No radioisotopes escape the containment vessels at the nuclear power plant.) The water passively shields the ADs from the walls, but radon is a radioactive inert gas and can find its way into the water through any number of pathways. Simple filtering will not remove it, although activated carbon has been shown~\cite{ISI:000286333400012} to be effective. However, the only isotope of radon that produces a significant background for the neutrino experiment, $^{222}$Rn, has a half-life of 3.8 days, rather less than the water residence time in the pools. Therefore our strategy for dealing with radon is to minimize its penetration into the water using a non-gas permeable pool liner and allowing any residual radon to simply decay away.

Temperature control was a design requirement, as the water pool must be held between $21^\circ$C and $25^\circ$C, with a target of $22.7\pm0.3^\circ$C. This is critical for the operation of the ADs, as they are filled with precise amounts of liquid scintillator mixtures and mineral oil, and are equipped with overflow tanks designed to handle the thermal expansion of the fluid within this band. An operating temperature of $22.7^\circ$C was chosen as it matched the rock temperature of the liquid scintillator production hall.

\section{System design and maintenance projections}

\begin{figure}[t]
\centering
\includegraphics[width=6.5in]{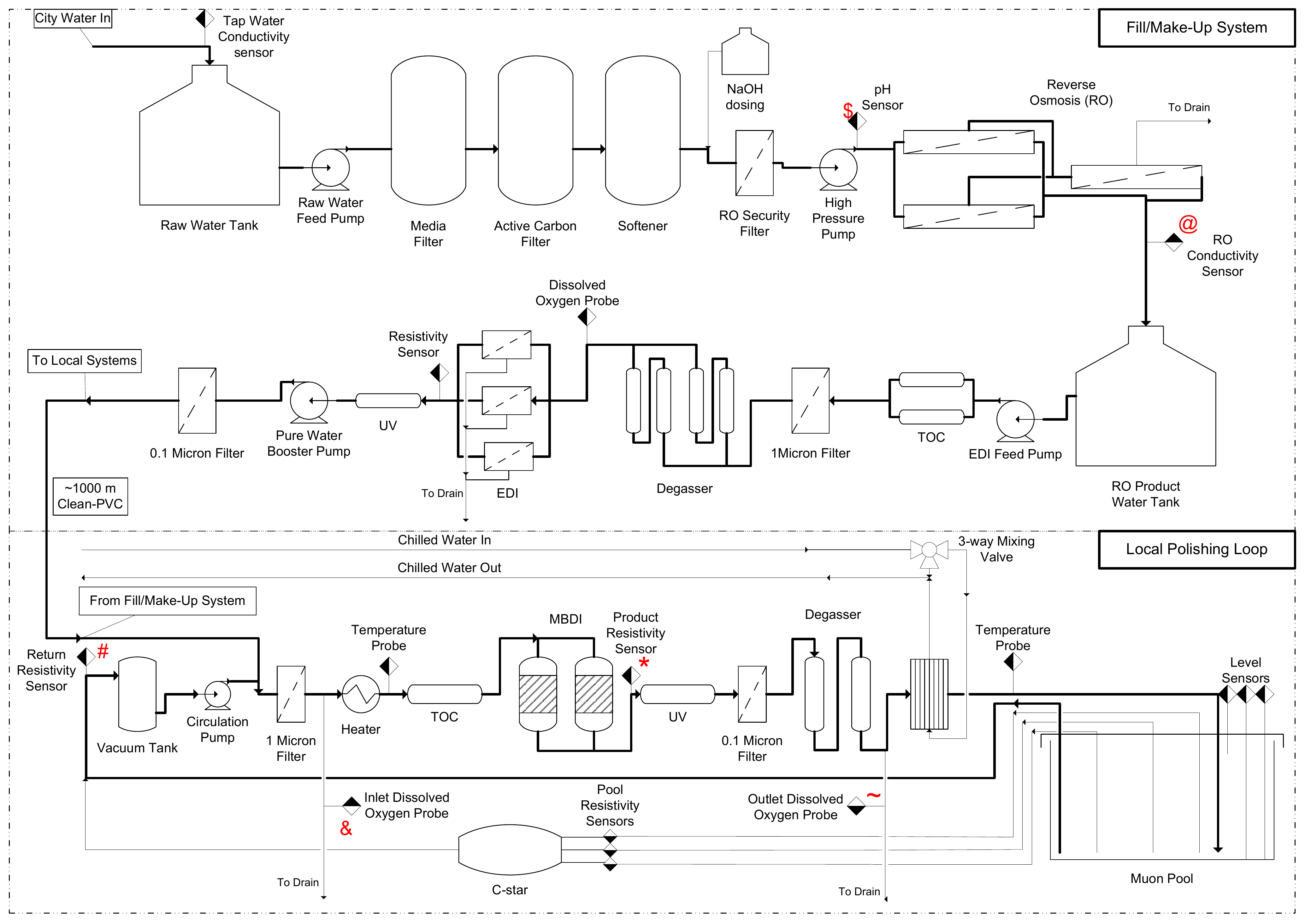}
\caption{Drawing of the central Fill/MakeUP (FMU) system feeding a local Polishing Loop (PL) system. Water resistivity, our key performance parameter, is measured at the output of the FMU, and both at the output and input of the PL. Note that the input to the PL is water from the pool during normal operation, but is from the FMU during filling periods. Various monitoring points are marked with symbols to which we refer in figures that follow.
\label{figure:WS}}
\end{figure}

An initial engineering design was drafted at Brookhaven National Laboratory (BNL). This design broke the system up into one Fill/MakeUp (FMU) stage that would feed each of the three pools, and individual Polishing Loop (PL) stages in each hall. An RFQ based on this design was sent to US and Chinese companies. We awarded the contract to Ultrapure Environmental Engineering (Ultrapure), a company based in Shenzhen, China, and the initial design was modified as the needs of the experiment became more clear. Figure~\ref{figure:WS}
shows the current layout of the FMU system feeding one of the three local PL systems.

The FMU is an 8 m$^3$/hr centralized treatment system to produce ultrapure water for filling and topping off of the pools. All water holding tanks in the FMU are polyethylene. The layers of the media filter are 400~kg of anthracite 0.8 to 1.8~mm, followed by 1250~kg of quartz sand 0.4 to 0.6~mm, then 195~kg of quartz sand 0.6 to 0.8~mm, and finally 195~kg of quartz sand 0.8 to 1.6~mm. The activated carbon filter is 400~kg of activated carbon, followed by 260~kg of quartz sand 0.8 to 1.6~mm, and then 150~kg of quartz sand 0.6 to 0.8~mm. The Reverse Osmosis (RO) security filters are 5$\mu$m meltblown polypropylene. Total organic carbon (TOC) removal is done via 185 nm UV lamp, and final deionization via electro-deionization units (EDI). Sterilization is accomplished with 254~nm UV lamps, followed by 1~$\mu$m and 0.1~$\mu$m filters made from folded membrane  polypropylene. A degassification stage is included just prior to the water entering the EDI.

The PL systems circulate at 5~m$^3$/hr or (for the large pool) 9~m$^3$/hr and are located in utility rooms in each experimental hall for maintaining water quality via continuous circulation.  The 1 and 0.1~$\mu$m filters are folded membrane  polypropylene. The vacuum tank was rubber lined fiber glass, but it was eventually replaced with stainless steel when ruptures seemed imminent. TOC removal is carried out via 185~nm UV, and mixed bed deionization (MBDI) maintains resistivity. Sterilization is done with 254~nm UV, and a stainless steel stacked plate cross flow heat exchanger (HX) maintains temperature. A degassification stage sits between the HX and the UV sterilizer. 

The FMU system is connected to the local PL systems via several hundred meters of Eslon\textsuperscript{\textregistered} Clean-PVC. Polyvinylidene fluoride (PVDF or Kynar\textsuperscript{\textregistered}) and other types of piping are better suited for use with high purity water, but the distances of piping needed ($\geq$5000 m) made the cost prohibitive. The entire system is controlled by a series of programable logic control (PLC) units interfaced with a central computer, providing data collection as well as operational control.

The FMU and PL systems use different deionization methods in the final stage. Electrodeionization (EDI) was chosen for the central processing system instead of mixed bed deionization (MBDI) to reduce the need for chemical expendables, as EDI can bring the RO product water to 18 M$\Omega$-cm continuously without needing replacement or regeneration and  the original design anticipated that the majority of ion removal would happen at this step of the treatment. Long term operations costs were considered in the choice of MBDI for the local recirculation systems, instead of EDI. Unlike EDI, MBDI can maintain water quality without electrical consumption beyond that of operating the pumps.

\subsection{Fill/Make Up (FMU) System}

The centralized FMU system consists of filtration, desalination, degassing, and sterilization elements. Municipally sourced water enters the system via a 4\inch~PVC line and flows into a 5~m$^{3}$  polyethylene (PE) equilization tank.
The source water then passes through one of two Grundfos CRN15-3 booster pumps, with AISI 316 Stainless Steel (SS) throughout, plumbed in parallel for redundancy via 2\inch~Chlorinated polyvinyl chloride (C-PVC) pipe at 12 m$^3$/hr. The water then passes through a 2 m$^3$ rubber-lined sand media filter tank constructed of carbon steel with an empty bed contact time (EBCT) of 6.8 minutes, and a 1.25~m$^3$ granulated active carbon filter tank of similar construction with an EBCT of 4.3 minutes, to be further refined; removing potential particulates, biological, and chemical contamination not removed during municipal treatment. The carbon is a coconut-based granulated charcoal manufactured by Hainan Xingguang Active Carbon Co., Ltd. 

The coarse filtered feed water passes through a 1.25~m$^3$ softening tank equiped, with Purolite C100E sulfonate cation exchange resin based on a crosslinked polystyrene gel bead, removing calcium and magnesium ions and preventing scaling on the RO membranes.  Next a sodium hydroxide dosing unit raises the pH of the water via injecting a 5-10\% NaOH solution, increasing the efficiency of the RO unit. This is followed by a stainless steel (SS) security filter housing fitted with seven meltblown polypropylene (PP) 30\inch~5$\mu$m filters to further decrease the turbidity of the water and to protect the RO membranes from any large particulates still in the feed water. The conditioned source water is then sent through a two stage RO unit by a 17 m$^{3}$/hr Grundfos CRCM15-12 high pressure pump with AISI 316 SS throughout.  The pretreated water is passed through two parallel Dunlop RO pressure vessels equipped with 4 Hydranautics CPA3-LD brine water RO membranes each. Product water from these parallel units is transferred to a 5 m$^{3}$ RO holding tank made of PE, while the reject water is passed through a final stage of RO before being piped into the holding tank. This two stage RO setup was desirable as it reduces concentrate water loses in the system to 23\%, a concern in system design. The product is a 9 m$^{3}$/hr flow rate of 1 to 0.1 M$\Omega$-cm water. After the RO unit all plumbing changes from C-PVC to Eslon\textsuperscript{\textregistered} Clean-PVC, chosen for its low cost to corrosion resistance ratio.

At this point the desalinated water passes through one of two parallel 9 m$^{3}$/hr Grundfos CRN10-6 booster pumps, with AISI 316 SS throughout, and begins the deionization, degassing, and sterilization processes necessary to produce the high quality water needed for DYB. The water passes through two parallel 300 Watt UV reactors to remove any organic carbon still present in the water. The units are manufactured by RenownUV and are comprised of a 316 SS flow cell with 185 nm UV lamps housed in quartz tubes. The units have a contact time of $\sim$9 seconds rated for 5 m$^3$/hr with 90\% TOC removal efficiency. This is then followed by a 304SS filter housing equipped with seven 30 inch 1 $\mu$m folded membrane PP filters, installed to ensure the stability and security of the input water quality to the downstream components. The water then enters four 6 X 28 Extra-flow Liqui-cel\textsuperscript{\textregistered} Membrane Contactor degassing units, plumbed in a two by two fashion (two units in series, two series in parallel).The membrane contactors contain several tightly grouped gas permeable tubes around which the water is forced by internal baffling. A single 1.5~hp Siemens pump provides vacuum for each degassing unit, which draws dissolved gasses out of the product water with a pressure differential of -0.1 MPa gauge. The degassing units remove $\sim$95\% of the dissolved gasses from the water, but yet another stage is needed to bring the water within the desired range. 

This is immediately followed by three IONPURE IP-LXM30Z EDI modules plumbed in parallel. The EDI runs continuously, producing water at $\geq$16 M$\Omega$-cm consistently with 90 to 95 \% permeate recovery. During a filling scenario, deionized water leaves the EDI unit and flows through a 150 W ultraviolet sterilizing unit with a 254nm wavelength constructed with 316 SS flow cell, designed for 9 m$^3$/hr and with a 4 second contact time, to mitigate bacteria growth in the processed water. The water then passes through one of two Grundfos booster pumps rated for 8 m$^3$/hr with AISI 316 SS throughout, before being forced through a 304SS filter housing with five 30\inch~0.1$\mu$m bonded PP membrane filters. At this point it is referred to as ultrapure water (UPW), and has a resistivity of approximately 18 M$\Omega$-cm. (This is consistent with ASTM Type~1 at this point, but as TOC, Na, Chlorides, Silica and Microbe counts are not specified, we did not confirm they are present to the ASTM specification.) If filling or topping is not taking place, deionized water is returned from the EDI to the RO holding tank and is run on a continuous loop to maintain quality.

\subsection{Localized Polishing Loop (PL) System}

Each Experimental Hall (EH)\footnote{The two smaller halls with neutrino detectors are called EH1 and EH2, and the large hall is EH3. For historical reasons, the grotto that houses the FMU is called EH4.} is equipped with a polishing system to further purify output water and to maintain water quality during circulation. Water entering the polishing system from the central system first passes through a 304SS pre-filter housing optimally equipped with five 30\inch~1$\mu$m PP membrane filters to protect downstream components from any large particulates picked up in the pool. This is then followed by the first stage of temperature control in the system, a 50kW  Berlin electric water heater, with 316SS internal plumbing. The unit is set to heat the water to 25$^o$C, slightly higher than the target temperature. It is then cooled to the target temperature in a stacked SS plate Heat Exchanger (HX). Although the source water is always too warm, this heating and cooling process provides a finer control over the temperature than cooling alone.

This is immediately followed by a 185 nm UV TOC removal device of the same construction and manufacturer as the ones used in the central system. This unit was installed to remove internal sources of organic matter. Water is then passed through 2 MBDI units containing 100 L of resins plumbed in parallel. The MBDI units are filled with DowEx$^{TM}$ mono-sphere MR-575LCNG resins, a 1:1 cation:anion resin of sulfonic acid and quaternary ammonium held in a styrene-divinylbenzene (DVB) matrix.

The MBDI is followed by a 254 nm UV sterilizer, identical to the unit in the fill make-up hall, and implemented to mitigate bacterial growth during recirculation. This is followed by a 304SS  filter housing equipped with five 20\inch~(EH1/EH2) or 30\inch~(EH3)~0.1$\mu$m PP membrane filters for removing sterilized bacteria, as well as removing any remaining particulates. Two more Liqui-cel\textsuperscript{\textregistered} degassing units in series follow this to further reduce dissolved gases in the water, especially CO$_2$, O$_2$, and Rn gases. The pools are optically and atmospherically sealed with a cover constructed of US~PTO Class~523 rubberized fabric with sulfur surface processing. The cover provides shielding from the ambient hall light so that the PMTs are able to detect the Cherenkov light produced by cosmic rays. Additionally, the cover helps maintain water quality by preventing particulates from entering the pool and limiting atmospheric gas dissolution. This is further enhanced by positively pressurizing the head space with N$_2$ gas from a 99.999\% pure liquid nitrogen boil off.

The water is run through the HX to precisely hold the final temperature to 22.7$^\circ$C before entering the pool. The HX is fed water from the experiment's chilled water supply via a three-way proportional valve adjusted using a feedback control loop. The water then flows into the pools and disperses through the inner and outer zones of the pool. The water is returned to the system via a triple--tube siphon, returning water separately from the inner pool, the outer pool, and the thin layer between the Tyvek\textsuperscript{\textregistered} liner and the pool wall.

The siphon is drawn by one of two Grundfos CRNCM10-4 circulation pumps, with AISI 316 SS throughout rated for 10 m$^3$/hr into a 0.2 m$^3$  vacuum tank with the inlet raised with respect to the outlet. These vacuum tanks were originally rubber lined fiber glass, but were eventually replaced with SS, for reasons described in Section~\ref{sec:failure}. The polishing system operates at 5 to 8~m$^{3}$/hr and is capable of producing 18.2~M$\Omega$-cm water with $\leq10$~ppb dissolved oxygen.

The system was sized to provide 8~m$^3$/hr during filling, and 5~m$^3$/hr and 9~m$^3$/hr, near and far site systems respectively, during circulation. Approximately, 3 to 4~m$^3$ of water are lost per day per experimental hall through the dissolved oxygen sensor flow cells, resulting in each pool being topped off once every 1 to 2 days.

\subsection{The automated control system}

The water systems are controlled by PLC units housed in each water utility room, and connected to a central computer via Ethernet. The systems operate primarily via simple feedback control loops connected to downstream sensors and monitors. In the central system, a municipal water buffer tank level sensor controls the electronic municipal water entry ball valve, creating a self-contained control loop. Similarly, the RO holding tank controls, at a base level, the operation of the all the components between itself and the municipal water equalization tank. Within this loop, a pH sensor controls the operation of the upstream NaOH dosing unit, regulating the pH of water prior to reaching the RO unit. The RO high pressure pump is controlled by pressure sensors upstream and downstream from its operation, allowing necessary pressures for RO operation to be generated, while avoiding damaging the membranes from excessive pressures. 

In each of the local systems, three level sensors monitor the pool water level. Two of these sensors provide level control, while the third is in reserve. In case one of the first two fail, this sensor will take its place as operation cannot cease to replace a sensor and access to the pool requires data taking to stop.
Temperature control within the systems is also managed via feedback control from temperature relay units placed after the heater and the HX.

This system not only controls daily operations of the water system, but provides initial data acquisition of many of the monitored parameters. This primary data acquisition system was interfaced with the experiments detector control system (DCS) to allow remote monitoring of the system's key parameters via internet. Parameters, such as pool outlet resistivity, product resistivity, inlet and outlet DO, and temperature can all be plotted in real time and are also archived. 

\subsection{System maintenance}

Maintenance guidelines for the water systems were included in the operations manual produced by the system contractor Ultrapure Environmental Systems. Additionally, a maintenance contract has been signed with Ultrapure for the duration of the experiment for the maintenance of the systems larger components and those requiring expert knowledge to maintain. This contract covers the chemical cleaning of the RO, EDI, and degassing membranes as well as the replacement of the MBDI resins in the local polishing loops. Ultrapure's maintenance projections and contract covers the following:

\begin{list}{$\bullet$}{}
\item Chemical cleaning of RO, EDI, and Degassing membranes every 6 months
\item Replacement of RO and  EDI membranes  and sand and charcoal filter media every 3 years
\item Changing of 5  $\mu$m RO  pretreatment filters every 15 to 30 days or when the pressure drop reaches 0.10 MPa
\item Changing of 1 $\mu$m EDI pretreatment filters every 3 to 6 months or when the pressure drop reaches 0.07 MPa
\item Changing of 0.1, 1 $\mu$m filters every 6 to 12 months or when the pressure drop reaches 0.07 MPa
\item Changing of the DowEX mono-sphere mix bed ion exchange resins every 6 to 12 months, or when the output drops below 14 M$\Omega$-cm
\item Cleaning and calibration of all probes, sensors, and switchs every 3 months
\item Monthly inspection of all systems, even if there is no indication of problems, to reduce the probability of unpredictable events.
\item UV system bulb cleaning, testing, and replacement as necessary
\item Emergency response within 24 hours
\item Providing a permanent or temporary solution within 14 days of incident, mitigating system down time. It was determined that the experiment veto can operate without water recirculation for this period of time with minimal ill effects. This is discussed further in Section~\ref{sec:performance}
\end{list}

\section{System Operation}

Operating procedures evolved as we gained more experience. Each pool has been completely filled twice and partially drained a number of times. Care was taken to prevent contamination of the pool's surfaces from dust and other construction debris. A partial fill and drain was executed prior to fully filling the pools in an effort to remove any settled particulates from the pool's lower surfaces. In spite of this, however, heavy particulate loads were experienced at startup. (Details on dealing with this are discussed in Section~\ref{sec:upgrades}.)

\subsection{Testing}

We verified the systems consistently produced water exceeding specifications before they were used to fill a pool. Additionally, each system was required to pass an operational readiness review performed by senior project management and engineering. As the FMU was installed, the contractor tested the system after the installation of each new component under manual operation. Once the FMU was installed, the automated system was tested and tuned to eliminate any bugs in the software.  We used the same process during the installation of each local PL. The units were either operated with water passing to drain or as a closed loop.

\subsection{Process performance}
\label{sec:performance}

\begin{figure}
\centering
\includegraphics[width=6.5in]{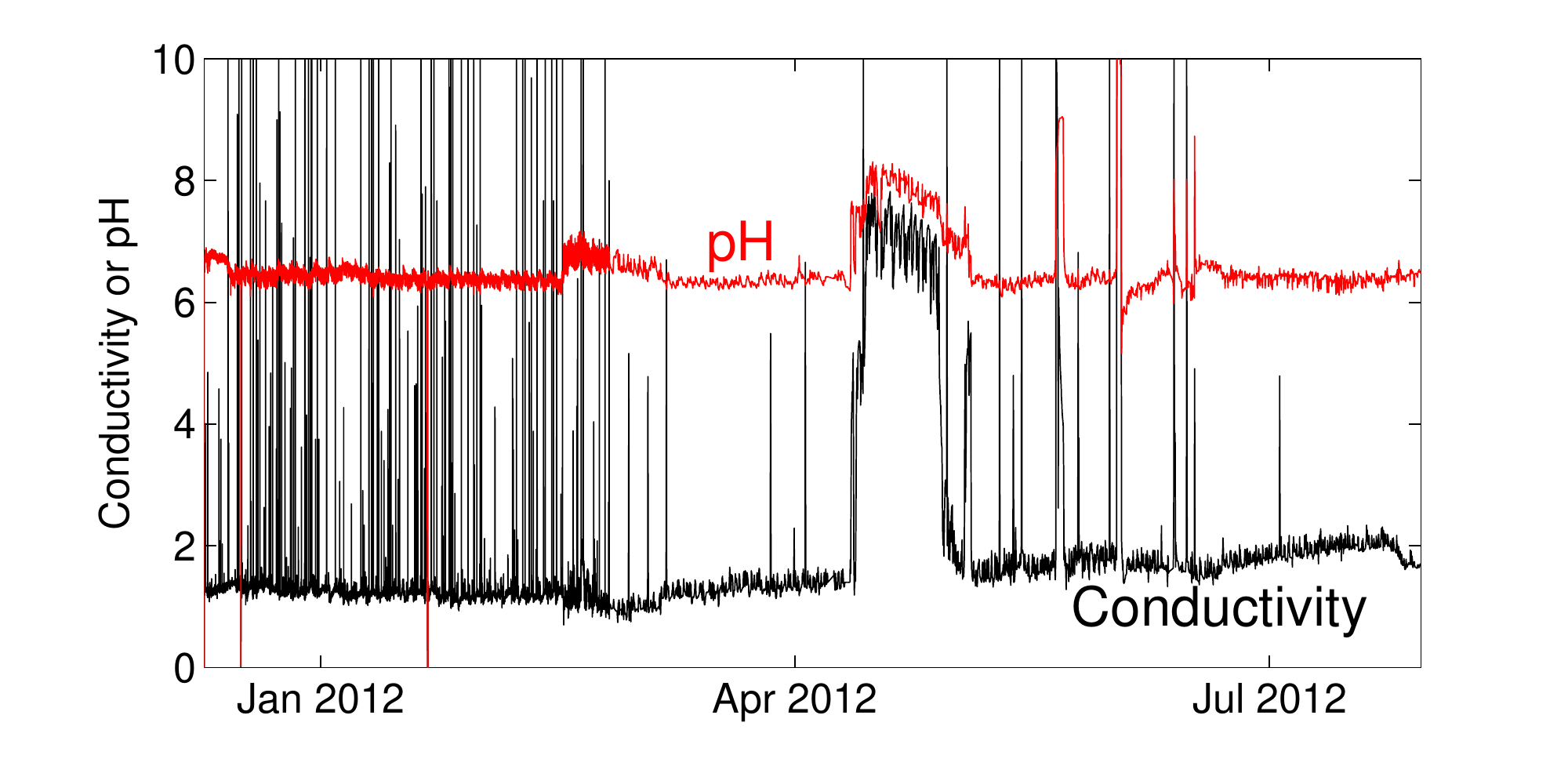}
\caption{RO product conductivity
(marked as {\color{red}@} in Fig.~\ref{figure:WS})
compared to pH ({\color{red}\$}) of dosed inlet water of the central FMU system, over several months. The vertical scale corresponds either to pH, or to conductivity in $\mu$S/cm.}
\label{figure:EH4RO}
\end{figure}

The central FMU system has operated with only minor tuning since April of 2011, consistently producing water with $\geq17$~M$\Omega$-cm resistivity and $\leq700$~ppb O$_2$ [aq] since the system was first fully operational. The high and consistent resistivity can be attributed to the effectiveness of the EDI, as well as the low conductivity produced by the RO. Figure~\ref{figure:EH4RO}
shows\footnote{
Plots such as Figure~\ref{figure:EH4RO} which show monitoring data over long periods of time, typically show a noisy ``grass'' where data values changed drastically over very brief intervals. This is because the sensors are located at junction points that sense different water quality when one of the three pools gets topped off with water from the FMU, and ultra pure water from the FMU passes through the local systems on its way to the pool, creating a distinct ``spike'' in the readings. Rather than artificially remove these spikes, we  emphasize that one should focus instead on the trends in the bulk of the data.}
how the RO removal efficiency is directly linked to the upstream pre-treatment. An unplanned increase in pH, in May~2012, directly leads to higher conductivity. 

\begin{figure}
\centering
\includegraphics[width=6.5in]{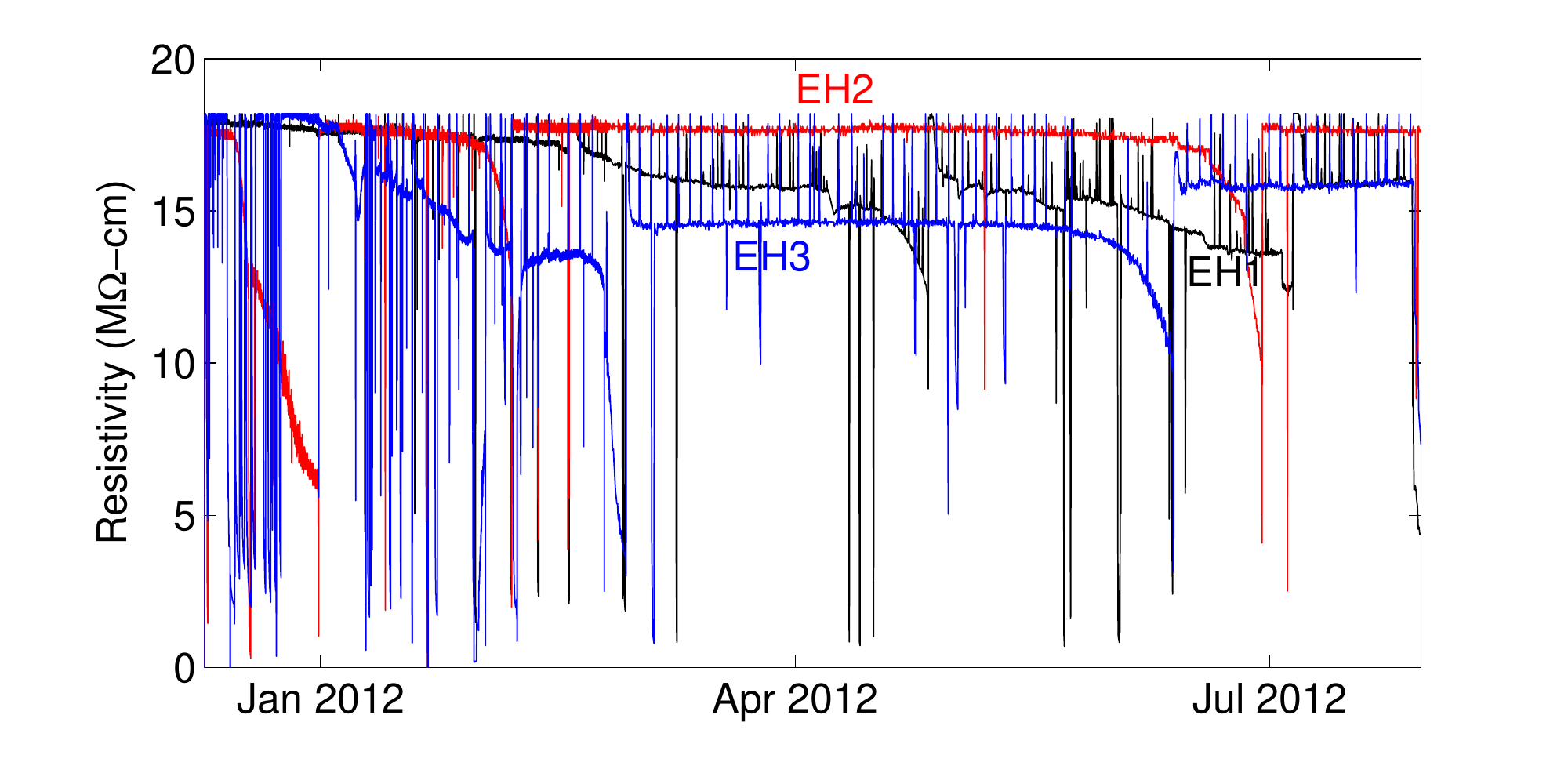}
\caption{Product resistivity
(marked as {\color{red}$\star$} in Fig.~\ref{figure:WS})
of each of the local PL systems (EH1, EH2, and EH3) from December 2011 to July 2012.}
\label{figure:ProductResist}
\end{figure}

The PL systems consistently produce water with $\geq14$~M$\Omega$-cm, except when the MBDI resins were allowed to degrade to observe the effect on pool water resistivity. Figure~\ref{figure:ProductResist}
illustrates the dramatic drop in product resistivity that follows. This was accomplished by consciously delaying the replacement of the MBDI resins until this signature asymptotic decline was observed in an effort understand mixing properties of the pool and the effect of varied input resistivity on veto performance and output resistivity.

\begin{figure}
\centering
\includegraphics[width=6.5in]{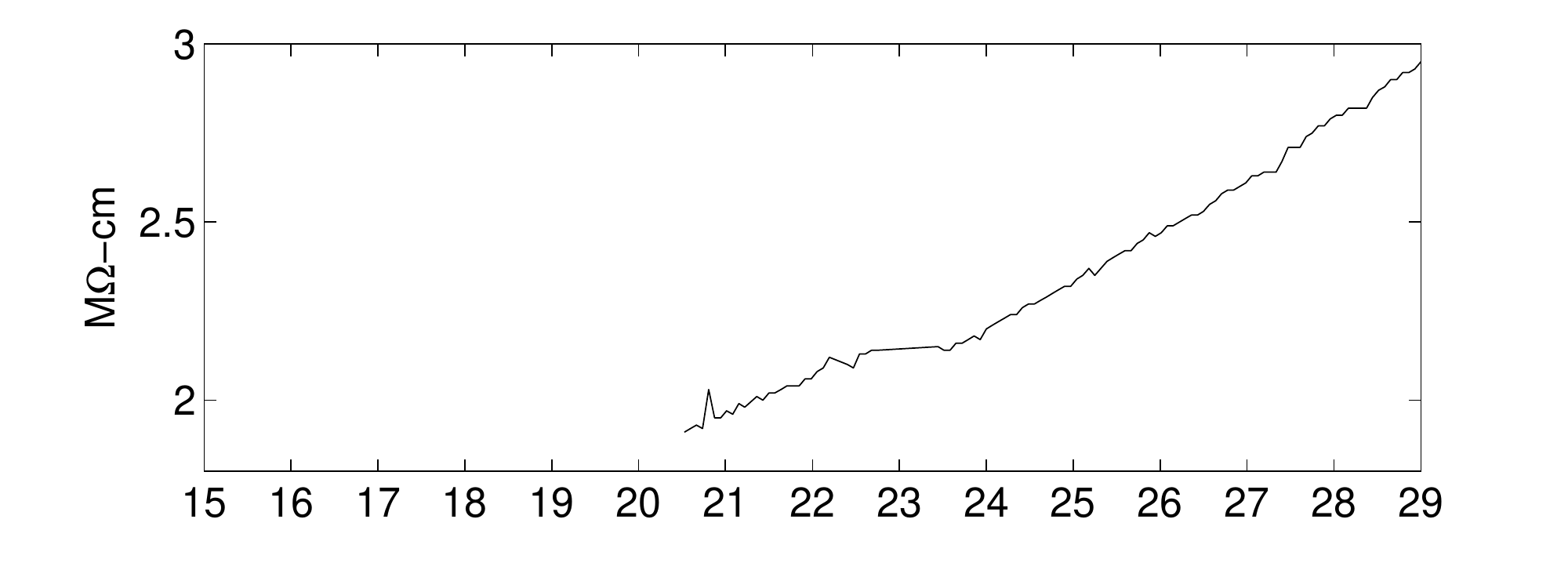}
\includegraphics[width=6.5in]{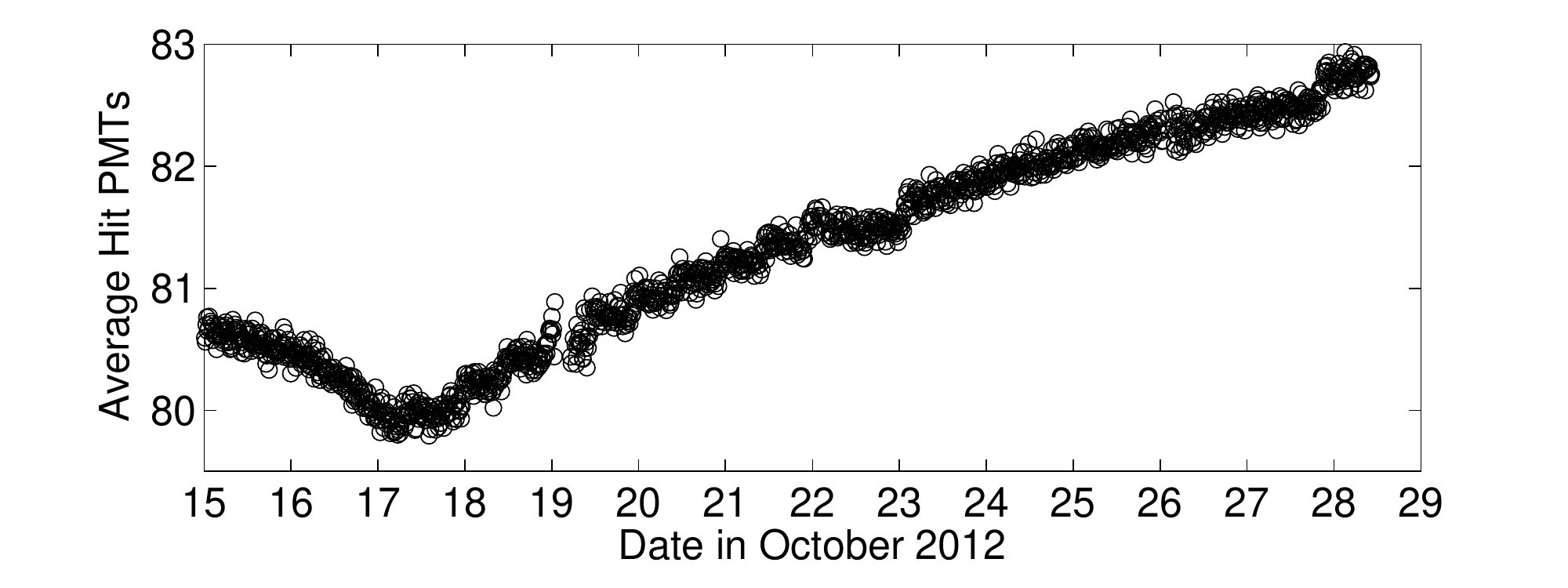}
\caption{EH2 pool resistivity (measured at the recirculation input to the PL,
marked as {\color{red}\#} in Fig.~\ref{figure:WS})
 and the average number of triggered photomultipliers per cosmic ray, during a period in October 2012, after repairs were made to the polishing loop. The correlation indicates an increasing attenuation length in the water with rise in resistivity.}
\label{figure:EH2Mounnhit}
\end{figure}

A clear effect on the average amount of detected light per cosmic ray event, can be seen when circulation is stopped for extended periods of time. For two weeks in October~2012, the EH2 local polishing system was non-operational. Figure~\ref{figure:EH2Mounnhit}
shows the effect during the one week period after repairs were completed. The rise in pool water resistivity from $\sim2$~M$\Omega$-cm to $\sim3$~M$\Omega$-cm correlates with a $1.5\%$ increase in the average number of signal PMTs per cosmic ray.
 
\subsection{Resistivity Changes and Contamination from Pool Surfaces}

\begin{figure}
\centering
\includegraphics[width=6.5in]{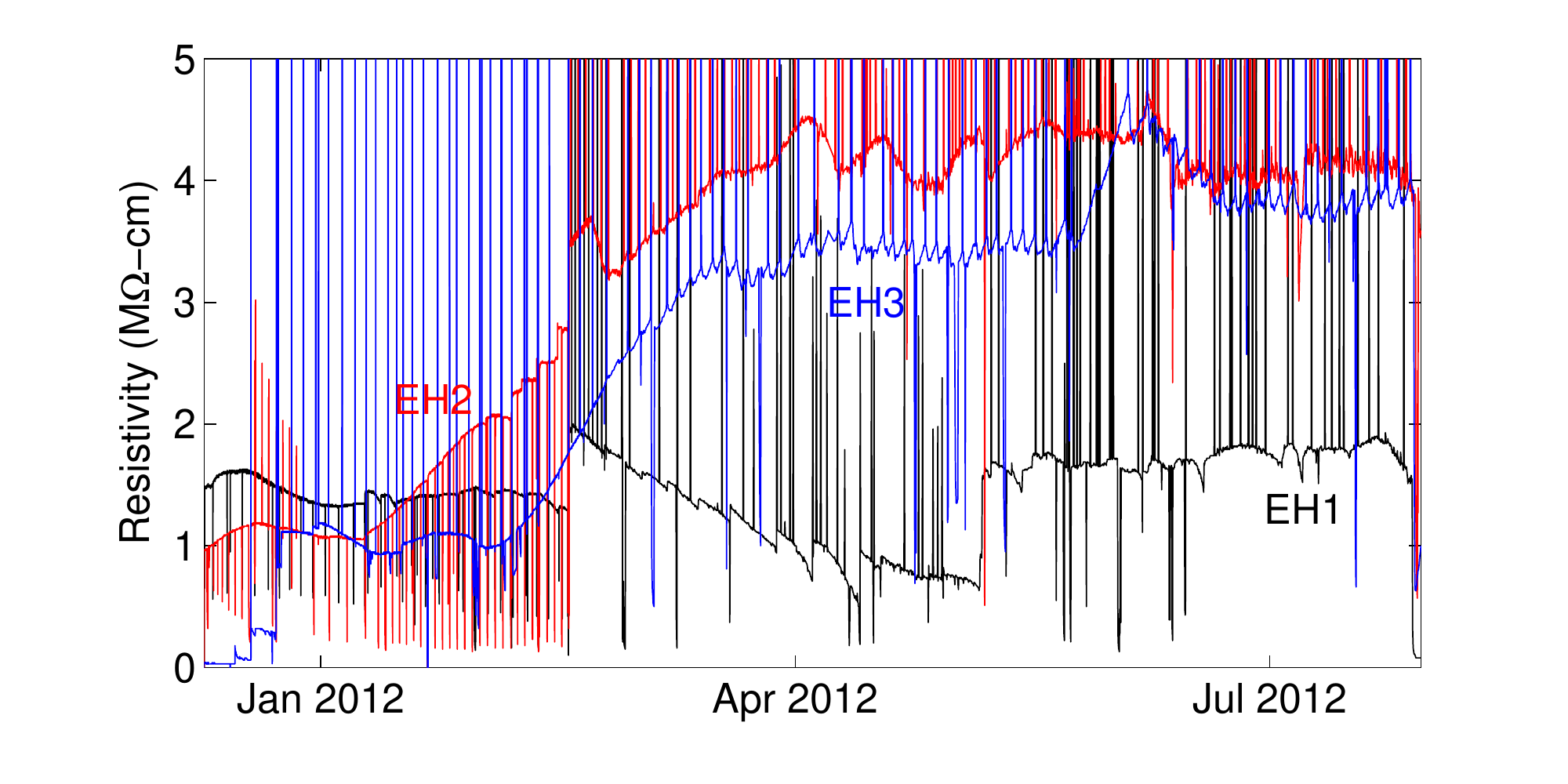}
\caption{Resistivity of water at the PL input
(marked as {\color{red}\#} in Fig.~\ref{figure:WS})
 for EH1, EH2, and EH3 for the same time period as shown in Fig.~\ref{figure:ProductResist}. This is nominally the pool water resistivity, except for the spikes during short periods of filling.}
\label{figure:PoolResist}
\end{figure}

The MBDI units raise the resistivity of the water entering the pools to 14~to 17~M$\Omega$-cm and the degassing units are capable of lowering dissolved oxygen (DO) concentrations to $\leq10$~ppb. However, water exiting the pools at the PL recirculation input has never been larger than $\sim8.5$~M$\Omega$-cm, as shown in Figure~\ref{figure:PoolResist}.
We presume that this reduction in resistivity comes from contamination contributed by surfaces in contact with the pools, although the mechanism for this is not yet understood. Possible sources include Fe ions leached from the stainless steel due to a corrosion mechanism; dissolution of CaCO$_3$ and SiO$_3$ from concrete and construction dust; carbon dioxide dissolution due to poor sealing of the pool cover, and bacterial metabolism supported by trace dissolved oxygen and dissolved organics leached from system components. All of these potential sources of contamination have been examined in an effort to determine their effect, and no single source can be identified as the primary reason for somewhat degraded water quality. 

Some clues to the nature of the contamination came accidentally. In early January 2012, the water quality had not improved after months of circulation, including the persistence of floating dust particles, although the cosmic ray veto performed adequately. It was discovered that a bypass valve was mistakenly left partially open  and only a sixth of the water was being treated. This bypass was closed and as these particles were removed, both the water quality and light collection improved. With time the resistivity of the water rose from $\leq1$~M$\Omega$-cm to $\sim2$~M$\Omega$-cm in EH1, and $\sim4$~M$\Omega$-cm in EH2/3. 

\subsubsection{Metalic organic complex}

\begin{figure}
\centering
\includegraphics[width=6.5in]{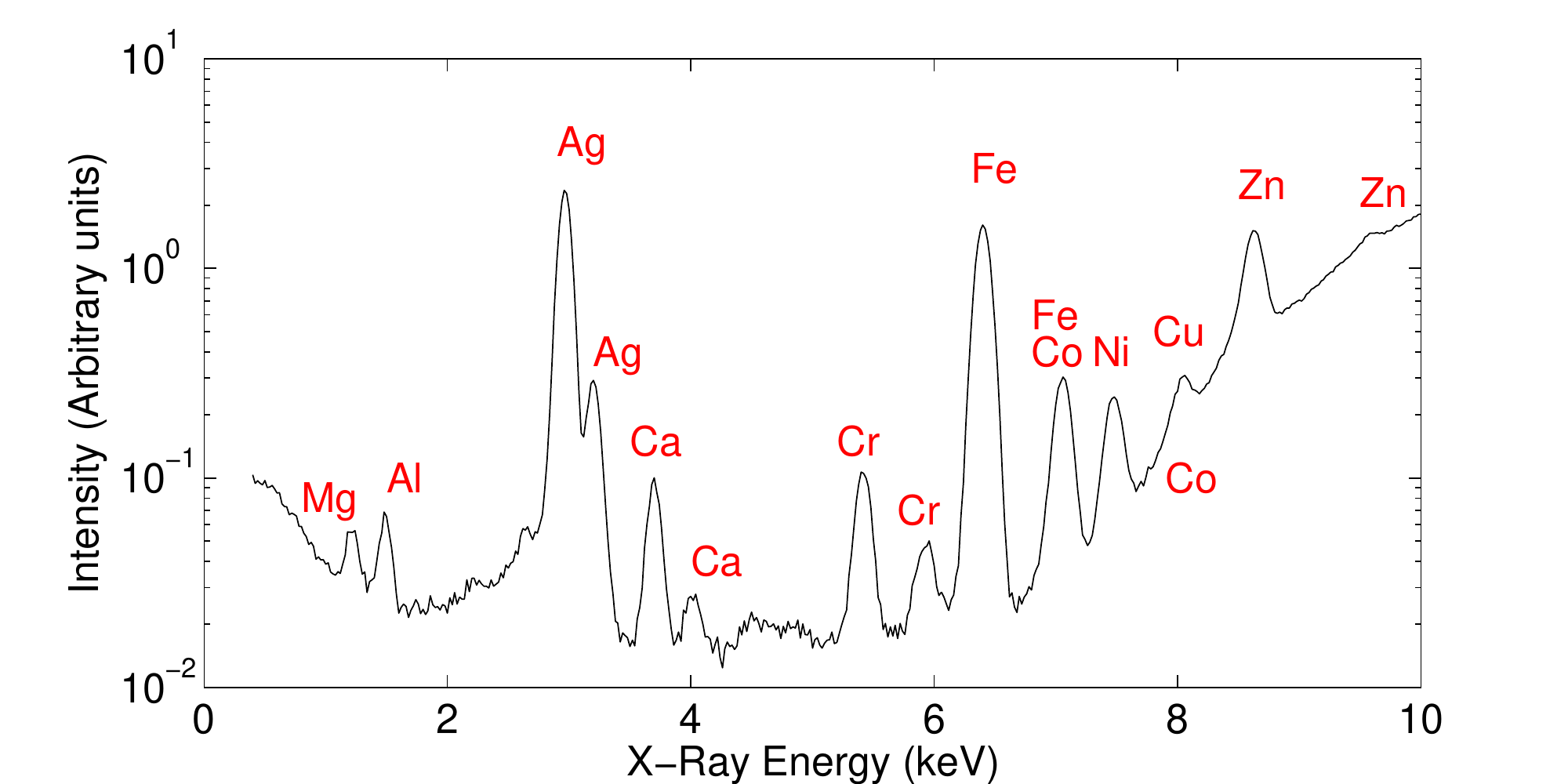}
\caption{XRF analysis of EH1 sludge sample taken March 2012.}
\label{figure:GoopXRF}
\end{figure}

In early March 2012, a buildup of yellow solids was discovered in the EH1 pre-filter housing. Approximately 10~mL of this material was removed from the housing and sent to BNL for analysis, along with sections of the filter from the housing. An X-ray Fluorescence (XRF) analysis, shown in Figure~\ref{figure:GoopXRF}, determined that the material was a metallic-organic complex containing Fe, Zn, and Ag compounds qualitatively.
Quantitative measurement of concentrations was not possible due to insufficient sample size. However, an outside laboratory determined that $\emph{Pseudomonas aeruginosa}$ was also present in the sample (contamination during gathering of the sample is suspected due to the prominence of this microbe in most environments).

It was decided that limiting sources of substrate and nutrients through continued treatment of the water was preferred to other methods of treatment that would interfere with the effectiveness of the veto. For Example, IMB\footnote{This was communicated to us as an internal report, from 1984, by Daniel Pope (Professor of Biology at Rensselaer Polytechnic Institute, and President of Biotest, Inc.) prepared for the IMB collaboration.} had limited long term success in treating Beggiatoa and Pseudomonas found in their system with peroxide. Subsequent samples taken after this indicated decreased bacterial levels.

\subsubsection{Carbon dioxide}

We took 59 water samples, from different pools and different locations, to analyze for CO$_2$. The samples were transported in biological oxygen demand (BOD) bottles, and analyzed at Rensselaer Polytechnic Institute via gas chromatography with methanization and flame ionization detection (FID).  Unfortunately, these samples failed to provide any conclusive results.  Measured Dissolved Inorganic Carbon (DIC) had significant variation within samples from the same sample port, and no conclusions or trends could be deduced from the data over all. This is most likely the result of a combination of low concentrations and imperfect sampling technique. Indeed, a theoretical calculation~\cite{CO2UPW} of the resistivity attributed solely to aqueous CO$_2$ at 22.5$^\circ$ suggests that concentrations as low as 10-100~ppb (by weight) would be enough to affect resistivity at our observed levels. Furthermore, this method does not distinguish between CO$_2$ from atmospheric dissolution and microbial metabolism. Both sources may have contributed to the concentrations necessary to lower the resistivity of the pools.

Table~\ref{tab:Auxres}
\begin{table}
\begin{center}
\caption{Data gathered from auxiliary resistivity sensors.}\label{tab:Auxres}
\begin{tabular}{|lrrrrrr|}
\hline
Date & Hall & Pool ``out" & RPI 1 & RPI 2 & BNL & IHEP\\
&  & M$\Omega$-cm & M$\Omega$-cm & M$\Omega$-cm & M$\Omega$-cm & M$\Omega$-cm \\
\hline
01/30/13 & EH1 & 1.40 & 5.08 & 5.09 &5.11 & $-$\\
01/31/13 & EH1 & 1.45 & $-$ & $-$ &$-$ & 5.25\\
03/12/13 & EH1 & 1.35 & $-$ & $-$ &1.35\footnotemark[1] & 1.27\footnotemark[1]\\
03/16/13 & EH1 & 1.50 & 6.09 & 6.09 &5.93 & $-$\\
\hline
03/11/13 & EH3 & 7.8 & 6.99 &6.89 &6.9 & 6.9\\
03/12/13 & EH3 & 8.0 & 6.09 & 6.09 &5.6\footnotemark[1] & 5.5\footnotemark[1]\\
03/19/13 & EH3 & 7.9 & 8.02 &7.7\footnotemark[2] &$-$ & $-$\\
\hline
03/13/13 & EH2 & 2.7\footnotemark[3] &3.14 & $-$ &$-$ & $-$\\
03/21/13 & EH2 & 1.21 &2.33 & $-$ &$-$ & $-$\\
03/22/13 & EH2 & 1.37 &2.8 & $-$ &$-$ & $-$\\
03/23/13 & EH2 & 1.55 &2.33 & $-$ &$-$ & $-$\\
03/26/13 & EH2 & 2.28 & 3.61 & $-$ &$-$ & $-$\\
03/27/13 & EH2 & 2.58 & 3.50 & $-$ &$-$ & $-$\\
03/28/13 & EH2 & 2.69 & 3.46 & $-$ &$-$ & $-$\\
\hline
\end{tabular}
\end{center}
{\footnotesize{
$^1$Cross check of pool outlet sensor.\\
$^2$System was stopped before a stable reading was made, and pool cover was removed.\\
$^3$System was not recirculating, last stable value was on Feb 28.}}
\end{table}
summarizes data gathered from additional sensors (detailed in section \ref{sec:upgrades}) that provides insight into the resistivity drop that seems to occur within the pools. There is a disparity between the resistivity of the water measured in the pool return line and that of the water in the pool center, a disparity which is quite large in the case of EH1.  We suspected that irregular flow and increased potential organics between the pool wall and the liner, a semi-isolated circulation zone that might allow sufficient microbial growth, could cause this disparity. Subtle differences in conditions at the pool walls could also account for the large differences between the different halls, as shown in Fig.~\ref{figure:PoolResist}.

To test the suspicion that the ultimate drop in resistivity came from microbe-laden pool walls, we shut down physics data taking in EH2 temporarily, and installed a ball valve to shut off water coming from the zone in-between the Tyvek\textsuperscript{\textregistered} lining and the pool wall. The resistivity from water returning from only the two pool sections was 3.5~M$\Omega$-cm, significantly higher than the 2~M$\Omega$-cm resistivity (at the time of this test) from all three return lines. This supports the idea that microbe growth on the pool wall is the source of the disparity.

\section{Multiple System Reproducibility and Process Modifications}

Though the local systems are identical in their components, order, and relative flow/volume ratios, a clear variation can be observed in the product resistivity (Figure~\ref{figure:ProductResist}) in each system, as well as the dissolved oxygen content and resistivity (Figure~\ref{figure:PoolResist}) of the water returning from the pools.There have been many suspected culprits investigated in an effort to determine not only why the water exiting the pools returns with such degraded quality, but what could be the sources of these system to system variations. In the process of these investigations, many bugs and minor problems have been discovered and corrected, though none could be held entirely responsible for the differences, or for the significant degradation of the input water.

\subsection{Equipment failure and replacement}
\label{sec:failure}
\begin{figure}
\centering
\includegraphics[width=6.5in]{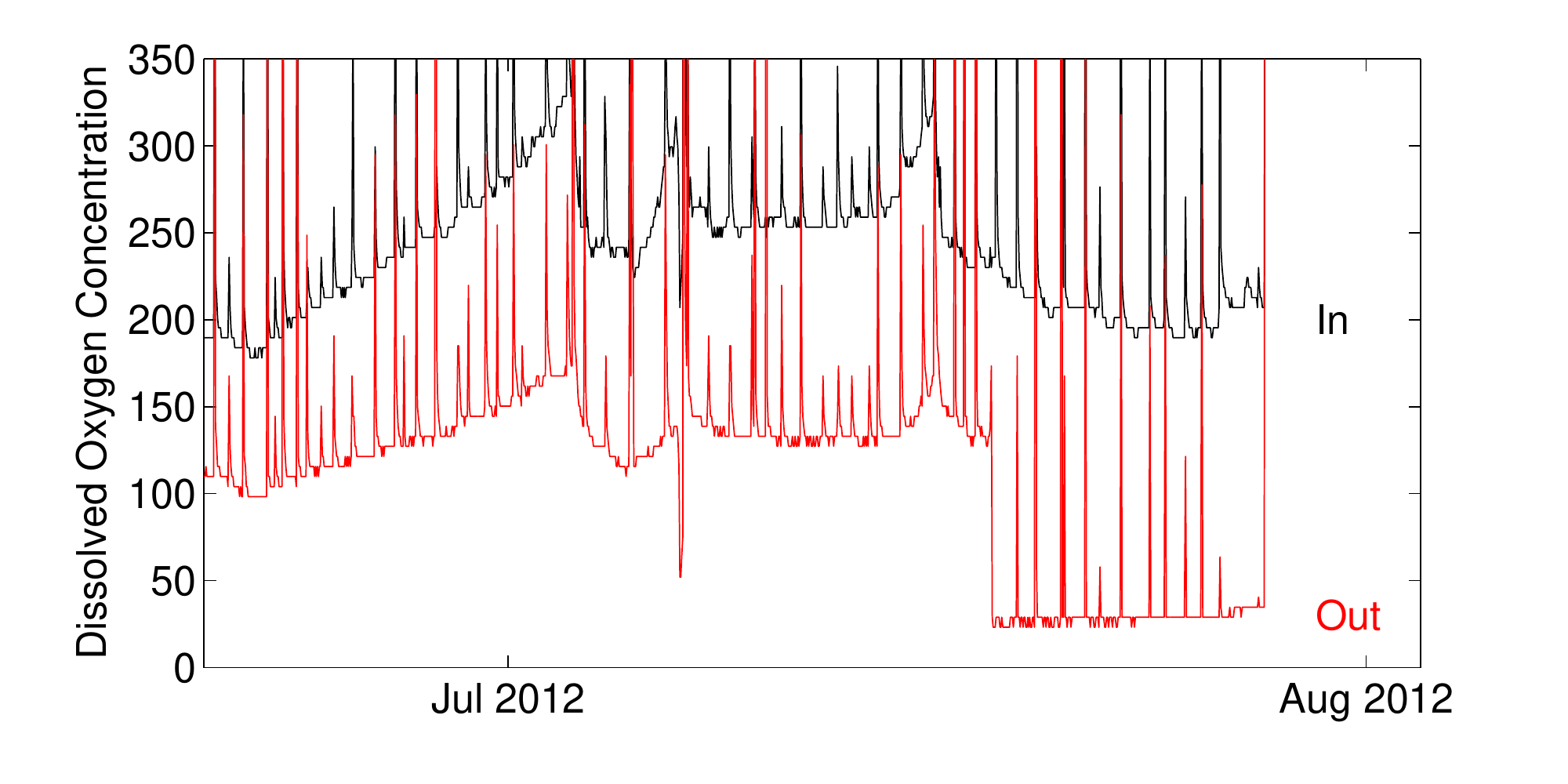}
\caption{Dissolved oxygen content (in ppb O$_2$) measured at the inlet
(marked as {\color{red}\&} in Fig.~\ref{figure:WS})
 and outlet
 ({\color{red}$\sim$}) points of the EH1 water system.}
\label{figure:EH1DOJuly}
\end{figure}

In June 2012, it was discovered that the vacuum pump for the EH1 degasser was not operating properly. Figure~\ref{figure:EH1DOJuly}
shows that this was corrected in mid-July, with a noticeable effect in the outlet DO concentration. However, the pool was uncovered and partially drained before the effect on the return water could be observed. After the EH1 degassing unit was repaired, its removal efficiency was on par or better than that of the other halls. 

During the run time of the experiment we have had a total of 4 level sensors fail in 3 separate halls. The reason for these failures is still unknown. The original system design called for 2 sensors per hall, one primary conductivity sensor and a secondary hydrostatic level sensor. In May of 2012, the hydrostatic level sensor in EH1 failed. Two hypothesis for the mode of failure are power surges/poor grounding and moisture entering the atmospheric vent tube. The decision was made to remove this sensor and install a replacement as well as additional hydrostatic sensors in each hall during a planned experiment-wide shutdown.

Shortly after the experiment came back on line, the level sensors in EH3 began failing. The conductivity level sensor began giving unreliable readings shortly after the water quality and resistivity in the pool began to rise. By January 21, 2013 all three level sensors had failed in this hall. After two months of estimation and manually filling the pool, five new sensors were installed. A 40 cm SS float sensor, a 50 cm conductivity level sensor, and 3 hydrostatic level sensors (two in reserve) with atmospheric venting tubes.It was believed that the hydro static level sensors failed because of humidity entering the atmospheric vent tube, which was located under the pool cover. In order to prevent this from happening again, the new hydrostatic sensors' vent tubes are located outside of the pool cover. It should be noted that because of this, the difference in pressure between inside the pool cover and outside can translate to up to 12 cm difference between measured and actual level.

On March 1, 2013, the EH2 vacuum tank split under the outward pressure of a leak check/filling. It was determined that the repeated pressurization of the vessel to check for leaks was the cause of the fatigue and eventual failure of the tank. New stainless steel tanks have been installed with additional pressure relief and gas purge valves to prevent future failures.

\subsection{Post installation upgrades}
\label{sec:upgrades}
In early December 2011, an additional 316SS filter housing was installed in the local polishing systems as a measure to protect downstream components from the previously mentioned high particle densities, and was equipped with 1 $\mu$m polypropylene membrane filters. While the addition of these filters were previously proposed during the design process by engineers at BNL (and a few companies that bid on the project) they were excluded from the contractors design and the system until four months after the water system was initially started. During this time the cost of changing several fouled resin beds and filters began to mount, and the installation cost of an upstream larger pore sized security filter became more favorable. 

In subsequent fillings, we decided to start with larger pore size filters in an attempt to avoid buildup of clogs, but still protect the MBDIs downstream. We used $2~\mu$m filters in the security filter housing and $1~\mu$m filters in the downstream housing. The system operated like this for several weeks until the filters ceased to experience significant pressure drops after several days of operation. At this point, we resumed the use of $1~\mu$m and $0.1~\mu$m filters.

After several months of working to diagnose the system behavior based on existing monitors, we decided to install additional sensors to directly monitor the water quality. Three lines were installed in each pool to take resistivity readings of the water directly with Rosemont Endurance 400 conductivity sensors with 0.01/cm cell constants. These lines combine into a single line housing a Western Environmental Technologies Laboratory's C-Star transmissometer~\cite{CStar} before entering the local circulation system at the vacuum tank. These lines operate using the same principle of the siphon return line used by the local system. Our first resistivity data is included in Table~\ref{tab:Auxres} under columns labeled ``RPI~1'' and ``RPI~2''. Our initial transmissometer measurements do indeed show that that (blue light wavelength) attenuation length is well above the 30~m specification.

\section{Conclusions and Recommendations}

We have demonstrated the ability, using only off-the-shelf technology, to build and operate a water purification system in a hostile environment, that provides several thousand m$^3$ of ultra pure water and reasonably maintains its purity through recirculation at several tons per hour. The resulting water has a clarity high enough to work very well as a cosmic ray veto system for a high energy physics neutrino experiment.

\begin{figure}
%\centerline{\includegraphics[width=5.0in]{VetoEfficiencyData.pdf}}
\centerline{\includegraphics[width=6.5in]{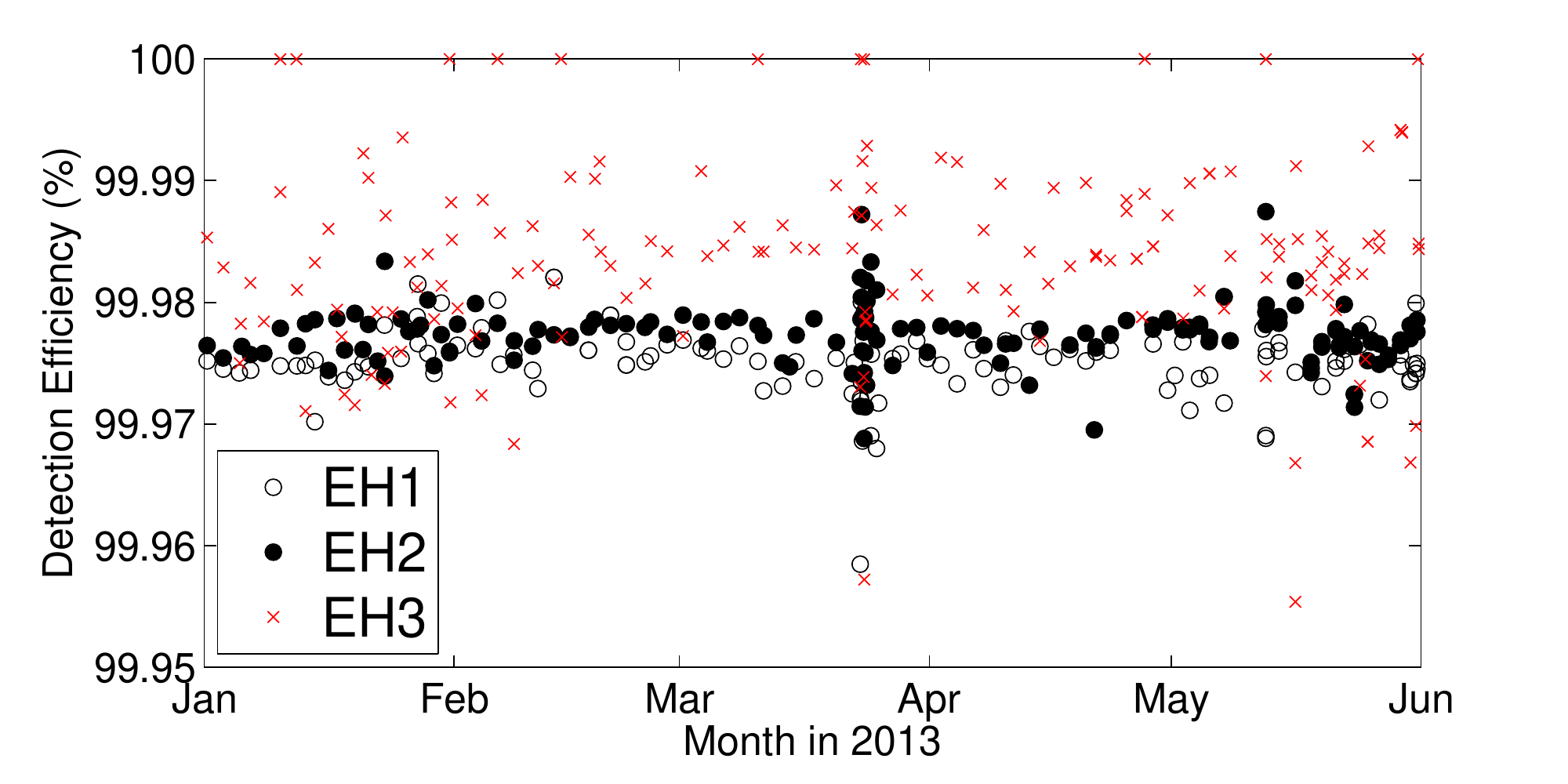}}
\caption{Measurements of the inner pool muon detection efficiency over the first half of 2013. Note that EH3 is deeper in the mountain, so fewer muons penetrate, leading to larger statistical scatter.
\label{fig:VetoPerformance}}
\end{figure}

Ultimately, the measure of our success is the long term efficiency of the water pool to actively reject cosmic ray muons. Figure~\ref{fig:VetoPerformance}
demonstrates an efficiency greater than 99.95\% over many months. Each AD can be used to identify a muon, and a corresponding signal is searched for in the water Cherenkov data. Results are consistent in all three pools, are constant over time, and this high veto efficiency has been a key ingredient to a successful reactor neutrino experiment.

The experiment, and the water system, are located in a relatively remote region in southeastern China. This necessitates a robust system that can operate with little intervention and infrequent active maintenance calls. At this time, the system has operated essentially continuously for more than two years, and is expected to continue operation for at least another three years.

Some questions remain, mainly as to the nature of the small resistivity changes that occur while the water is resident in the pool, between its production in the polishing loop and its withdraw back into this loop. Several potential sources of this resistivity change are offered, but as they do not significantly affect the operation of the experiment, these hypotheses have not been rigorously tested at this time. More monitoring points would be recommended for the next generation of such a system, and some way of continuously monitoring the water absorption length directly could be implemented.

\section{Acknowledgments}

The Daya Bay experiment is supported in part by the United States Department of Energy, office of High Energy Physics.

\section*{References}
\bibliographystyle{unsrt}
\bibliography{DayaBayWaterSystem}

\end{document}